# Modeling of Time with Metamaterials

Igor I. Smolyaninov, Yu-Ju Hung

*Department of Electrical and Computer Engineering, University of Maryland, College Park, MD 20742, USA*

**Metamaterials have been already used to model various exotic "optical spaces". Here we demonstrate that mapping of monochromatic extraordinary light distribution in a hyperbolic metamaterial along some spatial direction may model the "flow of time". This idea is illustrated in experiments performed with plasmonic hyperbolic metamaterials. Appearance of the "statistical arrow of time" is examined in an experimental scenario which emulates a Big Bang-like event.**

Nature of time has been a major subject of science, philosophy, and religion. Our everyday experiences tell us that time has a direction. On the other hand, most laws of physics appear to be symmetric with respect to time reversal. A few exceptions include the second law of thermodynamics, which states that entropy must increase over time, and the cosmological arrow of time, which points away from the Big Bang. While it is generally believed that the statistical and the cosmological arrows of time are connected, we cannot replay the Big Bang and prove this relationship in the experiment. Fortunately, it appears that electromagnetic metamaterials may provide us with interesting tools to better understand this relationship.

Recent advances in electromagnetic metamaterials already demonstrated highly unusual curvilinear "optical spaces" [1-3]. While nothing affects the flow of time in metamaterials, very recently it was demonstrated that a spatial coordinate may become "timelike" in a hyperbolic metamaterial [4]. To better understand this effect, let us start with a non-magnetic uniaxial anisotropic material with dielectric permittivities $\varepsilon_x = \varepsilon_y = \varepsilon_1$ and $\varepsilon_z = \varepsilon_2$, and assume that this behaviour holds in some frequency range around $\omega = \omega_0$. Any electromagnetic field propagating in this material can be expressed as a sum of the "ordinary" and "extraordinary" contributions, each of these being a sum of an arbitrary number of plane waves polarized in the "ordinary" ($\vec{E}$ perpendicular to the optical axis) and "extraordinary" ($\vec{E}$ parallel to the plane defined by the k–vector of the wave and the optical axis) directions. Let us define our "scalar" extraordinary wave function as $\varphi = E_z$ so that the ordinary portion of the electromagnetic field does not contribute to $\varphi$. This definition will be extremely important for the discussion of our results below, since it turns out that ordinary and extraordinary photons do not experience the same "effective metric". Since metamaterials generally exhibit high dispersion, let us work in the frequency domain and write the macroscopic Maxwell equations as

$$\frac{\omega^2}{c^2} \vec{D}_\omega = \vec{\nabla} \times \vec{\nabla} \times \vec{E}_\omega \ \text{ and } \ \vec{D}_\omega = \vec{\vec{\varepsilon}}_\omega \vec{E}_\omega \qquad (1)$$

Eq.(1) results in the following wave equation for $\varphi_\omega$ if $\varepsilon_1$ and $\varepsilon_2$ are kept constant inside the metamaterial:

$$-\frac{\omega^2}{c^2} \varphi_\omega = \frac{\partial^2 \varphi_\omega}{\varepsilon_1 \partial z^2} + \frac{1}{\varepsilon_2} \left( \frac{\partial^2 \varphi_\omega}{\partial x^2} + \frac{\partial^2 \varphi_\omega}{\partial y^2} \right) \qquad (2)$$

While in ordinary crystalline anisotropic media both $\varepsilon_1$ and $\varepsilon_2$ are positive, this is not necessarily the case in metamaterials. In hyperbolic metamaterials [5] $\varepsilon_1$ and $\varepsilon_2$ have opposite signs. These metamaterials are typically composed of multilayer metal-dielectric or metal wire array structures, as shown in Fig.1. Optical properties of such metamaterials are quite extraordinary. For example, there is no usual diffraction limit in a hyperbolic metamaterial [6,7]. Let us consider the case of constant $\varepsilon_1 > 0$ and $\varepsilon_2 < 0$ and assume that this behavior holds in some frequency range around $\omega = \omega_0$. Let us assume that the metamaterial is illuminated by coherent CW laser field at frequency $\omega_0$, and we study spatial distribution of the extraordinary field $\varphi_\omega$ at this frequency. Under these assumptions equation (3) can be re-written in the form of 3D Klein-Gordon equation describing a massive scalar $\varphi_\omega$ field:

$$-\frac{\partial^2 \varphi_\omega}{\varepsilon_1 \partial z^2} + \frac{1}{|\varepsilon_2|}\left(\frac{\partial^2 \varphi_\omega}{\partial x^2} + \frac{\partial^2 \varphi_\omega}{\partial y^2}\right) = \frac{\omega_0^2}{c^2}\varphi_\omega = \frac{m^{*2} c^2}{\hbar^2}\varphi_\omega \qquad (3)$$

in which the spatial coordinate $z = \tau$ behaves as a "timelike" variable. Therefore, eq.(3) describes world lines of massive particles which propagate in a flat (2+1) Minkowski spacetime (Fig.2d). When a metamaterial is built and illuminated with a coherent extraordinary CW laser beam, the stationary pattern of light propagation inside the metamaterial represents a complete "history" of a toy (2+1) dimensional spacetime populated with particles of mass $m^*$. This "history" is written as a collection of particle world lines along the "timelike" $z$ coordinate.

The world lines of particles described by eq.(3) are straight lines. These straight world lines are easy to observe in the experiment, as demonstrated in Fig.2. These experiments were performed using PMMA-based plasmonic hyperbolic metamaterials described in detail in ref. [6]. Rigorous theoretical description of these metamaterials has been developed in ref. [8]. While simple demonstration of a single straight "world line" presented in Fig.2 is almost trivial, situation may become much more interesting if we allow adiabatic variations of $\varepsilon_1$ and $\varepsilon_2$ inside the metamaterial. The Klein-Gordon equation for a massive particle in a gravitational field can be written as [9]:

$$\frac{1}{\sqrt{-g}} \frac{\partial}{\partial x^i} \left( g^{ik} \sqrt{-g} \frac{\partial \varphi}{\partial x^k} \right) = \frac{m^2 c^2}{\hbar^2} \varphi \qquad (4)$$

If $\varepsilon_1$ and $\varepsilon_2$ are allowed to vary, eq.(3) will remain approximately valid if $\partial \varepsilon / \partial x^i << k_i$. Comparison of eqs.(3) and (4) demonstrates that world lines of massive particles in some well known curvilinear spacetimes can indeed be emulated using hyperbolic metamaterials in the case of slowly varying $\varepsilon_1$ and $\varepsilon_2$. For example, let us consider an experimental situation, in which we allow slow adiabatic variation of $\varepsilon_2$ as a function of $z$, while $\varepsilon_1$ is kept constant. According to eqs.(3,4) this situation corresponds to "cosmological expansion" of the (2+1) dimensional universe as a function of "timelike" variable $z$. Therefore, spatial separation of particle world lines must increase with increasing $z$ (see Fig.3d). Some complications within such a model may arise due to loss and spatial dispersion in practical hyperbolic metamaterials. On the other hand, losses may be compensated if a gain medium is used as a dielectric component of the metamaterial, while effects of spatial dispersion are widely believe to affect the behavior of real spacetime at the Planck scale. Thus, difficulties presented to our model by losses and spatial dispersion are not insurmountable.

While experimental demonstration of the "expanding universe" with 3D metamaterials would require sophisticated nanofabrication, experimental demonstration of this concept using plasmonic hyperbolic metamaterials is much simpler. Let us start by re-writing eq.(1) in cylindrical coordinates $(r, \theta, z)$. Let us assume that $\varepsilon_\theta$, $\varepsilon_z$, and $\varepsilon_r$ change very slow as a function of coordinates so that their derivatives can be neglected.

In addition, let us assume that $\varepsilon_\theta = \varepsilon_z > 0$, $\varepsilon_r < 0$, and re-define our "scalar" extraordinary wave function as $\varphi = E_r$. Under these approximations we obtain

$$-\frac{\partial^2 E_r}{\varepsilon_\theta \partial r^2} + \frac{1}{|\varepsilon_r|}\left(\frac{\partial^2 E_r}{\partial z^2} + \frac{\partial^2 E_r}{r^2 \partial \theta^2}\right) - \frac{\partial E_r}{\varepsilon_\theta r \partial r} + \frac{E_r}{\varepsilon_\theta r^2} + \frac{\partial E_\theta}{\varepsilon_r r^2 \partial \theta} = -\frac{\partial^2 E_r}{\partial t^2} \qquad (5)$$

At large enough $r$ extraordinary rays become well defined, the last three terms on the left side of eq.(5) can be neglected, and the new scalar wave equation can be written [5] as

$$-\frac{\partial^2 \varphi_\omega}{\varepsilon_\theta \partial r^2} + \frac{1}{|\varepsilon_r|}\left(\frac{\partial^2 \varphi_\omega}{\partial z^2} + \frac{\partial^2 \varphi_\omega}{r^2 \partial \theta^2}\right) = \frac{\omega_0^2}{c^2}\varphi_\omega = \frac{m^{*2} c^2}{\hbar^2}\varphi_\omega \quad, \qquad (6)$$

where $r = \tau$ becomes a "timelike" variable (note that the local metric experienced by so-defined extraordinary field coincides with metric (3) at large enough $r$, while ordinary rays do not experience this metric).

In the case of plasmonic hyperbolic metamaterial eq.(6) can be simplified even further:

$$-\frac{\partial^2 \varphi_\omega}{\varepsilon_\theta \partial r^2} + \frac{1}{|\varepsilon_r|}\frac{\partial^2 \varphi_\omega}{r^2 \partial \theta^2} = \frac{m^{*2} c^2}{\hbar^2}\varphi_\omega \qquad (7)$$

Now it is apparent that the "cosmological expansion" as a function of $r = \tau$ may be realized with constant $\varepsilon_\theta > 0$ and $\varepsilon_r < 0$, which validates our original assumptions. Experimental demonstration of the world line behavior in an "expanding universe" as a function of "timelike" $r = \tau$ is presented in Fig.3. This experiment is similar to the experimental demonstration of the plasmonic hyperlens [6]. Plasmon rays are launched into the hyperbolic metamaterial near $r = 0$ point via the central phase matching structure marked with an arrow in Figs.3(a,b). Similar to the world line behavior near the Big Bang which is shown in Fig.3(d), plasmonic rays or "world lines" indeed increase their spatial separation as a function of "timelike" radial coordinate $r=\tau$. The point (or moment) $r=\tau=0$ corresponds to a moment of the toy "big bang". Comparison of Fig.3c with Fig.3d indicates that the "cosmological arrow of time" is well defined and points from $r=0$ outwards.

Let us examine if this experimental model may help us illustrate the relationship between the statistical and the cosmological arrows of time. As a first step, let us define the operational notion of entropy within the scope of our model. Let us follow the classical particle-in-a-box approach. Let us define our "boxes" as circular sectors with an angular width equal to the width of an individual plasmonic ray or "world line" in Fig.3c, and number these boxes from 1 to $m$. As a result, field intensity $I(r)$ at a given radius $r$ can be interpreted as a distribution of identical particles of mass $m^*$ among these boxes. Therefore, entropy can be written as

$$S = k \ln G = k \ln \frac{N!}{N_1! N_2! ... N_m!} \qquad (8)$$

where $N_m$ is the number of particles in box $m$, and $N$ is the total number of particles [10]. Since the numbers of particles are very large, we may use the Stirling's approximation: $\ln N! \approx N \ln N$ and obtain an approximate expression

$$S \sim I \ln I - \sum I_m \ln I_m \qquad (9)$$

where $I = \sum I_m$. Equations (8,9) allow us to calculate entropy $S(r)$ at a given radius or "moment of time" $r=\tau$, and examine if the "statistical arrow of time" in our model coincides with the "cosmological" one. The possible sources of entropy in these model experiments are nonlinear optical interactions of plasmonic rays and random disorder of the samples. In the absence of both factors "entropy" defined by eqs.(8,9) should remain constant. Ideally, the best case scenario for these experiments would be to use a perfect defectless hyperbolic metamaterial structure, which is made using a highly nonlinear dielectric, such as photorefractive lithium niobate or barium titanate. Such an experimental geometry would present us with a good opportunity to study entropy increase due to particle interaction in a truly irreversible nonlinear optical scenario. Unfortunately, in our current experiments we must rely only on the random disorder of the samples. This is analogous to experiments performed with a small number of molecules of an ideal gas, which is composed of perfectly elastic molecules. Entropy increase in such a gas is known to be reversible in principle. In a similar fashion, the effects of linear random scattering of light are known to be reversible using, for instance, adaptive optics approach. Therefore, our experiments must be considered only as an indication of potentially interesting research direction. The summary of these experiments is presented in Fig.4. In one of these experiments shown in Fig.4c, a low entropy state created at the moment of "big bang" by launching two plasmon rays into the structure (Fig.4d) has been gradually scattered by random defects into a high entropy state (Fig.4e). This behavior is strikingly different from the almost ideal defectless case presented in Fig3c, in which entropy stays almost constant. Fig.4f clearly demonstrates that entropy plotted as a function of "timelike" radial coordinate does grow with time in the presence of disorder. Thus, "statistical" and "cosmological" arrows of time do coincide in our model experiments.

Finally, let us ask a natural question if closed timelike curves (CTC) are allowed in a 3D hyperbolic metamaterial. At first glance, this question is simple, and the answer should be "yes". If the angular coordinate $\theta$ is made "timelike", while $r$ and $z$ coordinates are spacelike, we seem to have a clear case of CTC in which all the "grandfather paradoxes" are resolved by the cyclic boundary conditions. However, more detailed analysis described below indicates that Nature resists creation of even such trivial CTCs in a 3D metamaterial. In order to perform this analysis we need to assume that $\varepsilon_\theta$, $\varepsilon_z$, and $\varepsilon_r$ change very slow as a function of coordinates so that their derivatives can be neglected. In addition, let us assume that $\varepsilon_r = \varepsilon_z > 0$, $\varepsilon_\theta < 0$, and re-define our "scalar" extraordinary wave function as $\varphi = E_\theta$. Under these approximations we obtain

$$-\frac{\partial^2 E_\theta}{\varepsilon_r r^2 \partial \theta^2} + \frac{1}{|\varepsilon_\theta|}\left(\frac{\partial^2 E_\theta}{\partial z^2} + \frac{\partial^2 E_\theta}{\partial r^2}\right) - \frac{2\partial E_r}{\varepsilon_\theta r^2 \partial \theta} = -\frac{\partial^2 E_\theta}{\partial t^2} \qquad (10)$$

At large enough $r$ extraordinary rays become well defined, the last term on the left side of eq.(10) can be neglected, and the new scalar wave equation can be written as

$$-\frac{\partial^2 \varphi_\omega}{\varepsilon_r r^2 \partial \theta^2} + \frac{1}{|\varepsilon_\theta|}\left(\frac{\partial^2 \varphi_\omega}{\partial z^2} + \frac{\partial^2 \varphi_\omega}{\partial r^2}\right) = \frac{\omega_0^2}{c^2}\varphi_\omega = \frac{m^{*2}c^2}{\hbar^2}\varphi_\omega \quad , \qquad (11)$$

where $\theta = \tau$ becomes a "timelike" variable (once again, note that the local metric experienced by so-defined extraordinary field coincides with metric (3) at large enough $r$, while ordinary rays do not experience this metric). An infinite number of CTCs defined by condition $r = const$ and $z = const$ do appear in such a toy spacetime, and an extraordinary ray may be sent over a "world line", which is very close to a CTC. However, the only rays which can propagate along exact cyclic CTC trajectories are ordinary rays. These rays do not experience effective metric defined by eq.(11), and they do not perceive $\theta$ as a timelike variable.

**Figure Captions**

**Figure 1.** Schematic views of "wired" (a) and "layered" (b) hyperbolic metamaterials. (c) Hyperbolic dispersion relation is illustrated as a surface of constant frequency in k-space. When $z$ coordinate is "timelike", $k_z$ behaves as effective "energy".

**Figure 2.** Experimental demonstration of straight "world lines" in a hyperbolic metamaterial: (a) Image of the plasmonic hyperbolic metamaterial obtained using optical microscope under white light illumination. The defect used as a plasmon source is shown by an arrow. (b) AFM image of the metamaterial shows stripes of PMMA formed on the gold film surface using E-beam lithography. (c) Straight plasmonic ray or "world line" is emitted from the defect under illumination with 532 nm laser light. The ray direction is indicated by the arrow. For the sake of clarity, light scattering by the edges of the PMMA pattern is partially blocked by semi-transparent rectangles. (d) Schematic view of a particle world line in a (2+1) dimensional Minkowski spacetime.

**Figure 3**. Experimental demonstration of world line behavior in an "expanding universe" using a plasmonic hyperbolic metamaterial: Optical (a) and AFM (b) images of the plasmonic hyperbolic metamaterial based on PMMA stripes on gold. The defect used as a plasmon source is shown by an arrow. (c) Plasmonic rays or "world lines" increase their spatial separation as a function of "timelike" radial coordinate. The point (or moment) $r=\tau=0$ corresponds to a toy "big bang". For the sake of clarity, light scattering by the edges of the PMMA pattern is partially blocked by semi-transparent triangles. (d) Schematic view of world lines behavior near the Big Bang.

**Figure 4**. (a,b,c) Effect of various degree of disorder on field distribution inside the metamaterial structure shown in Fig.3. The cross section of image (c) near the launch point (d) and near the outer rim of the structure (e) demonstrates increase of entropy. (f)

Entropy (as defined by eqs.(8,9)) plotted as a function of "timelike" radial coordinate for the experiment presented in (c) demonstrates that "statistical" and "cosmological" arrows of time do coincide in our experiments.

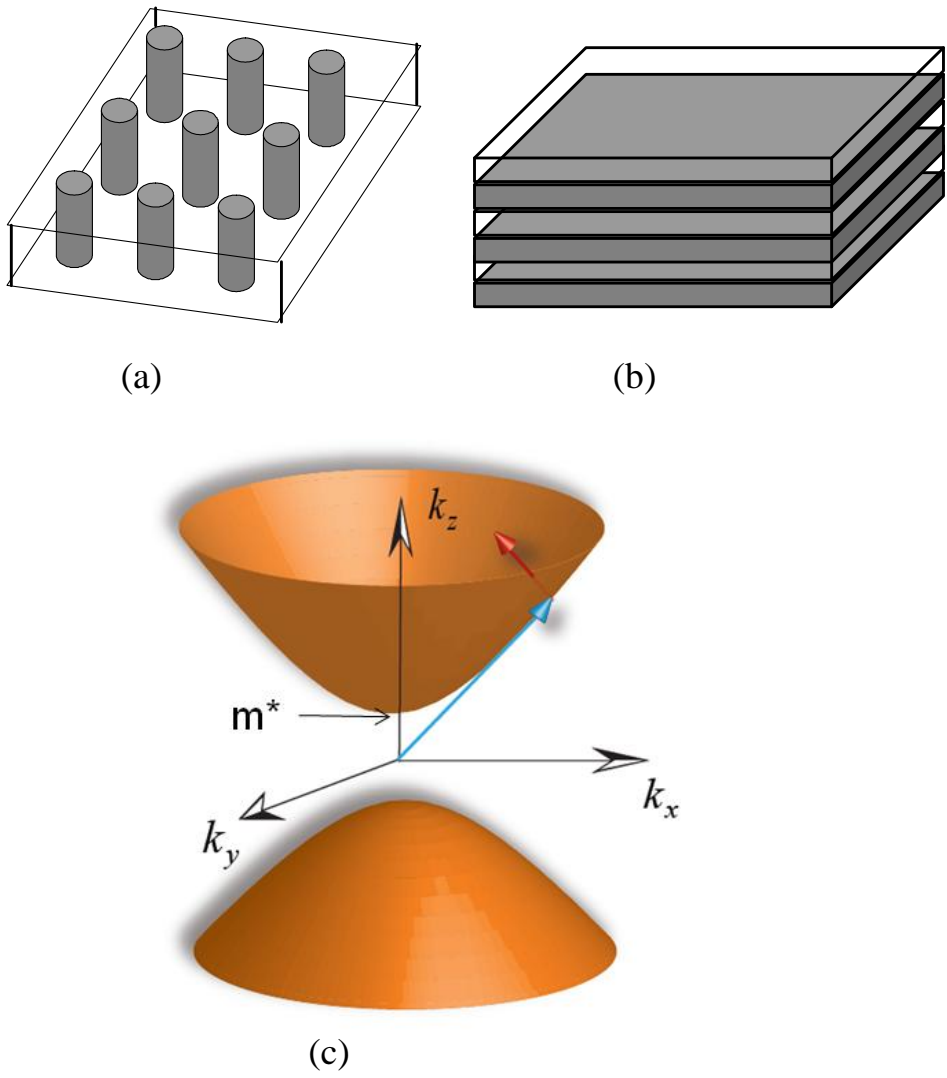

(a)

(b)

(c)

Fig.1

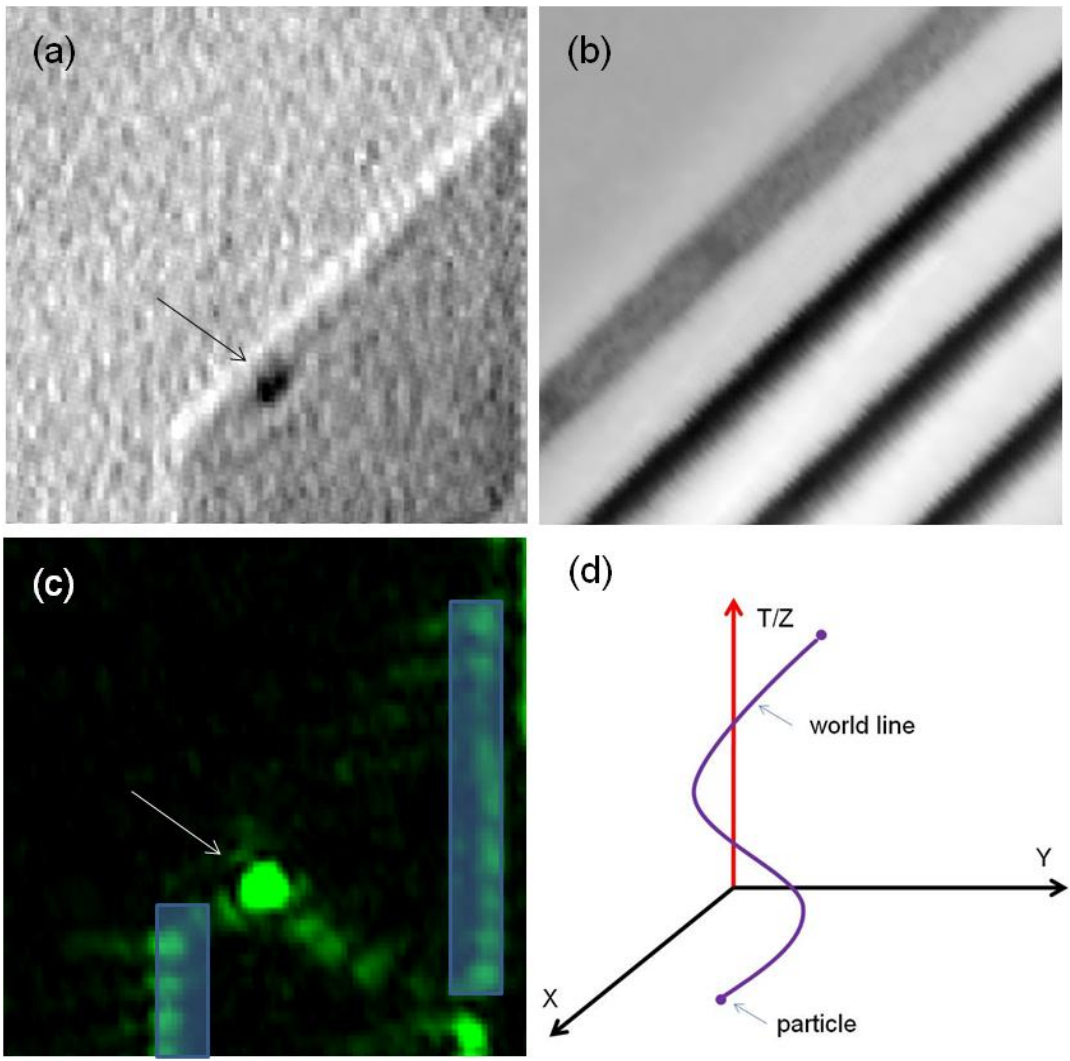

Fig.2

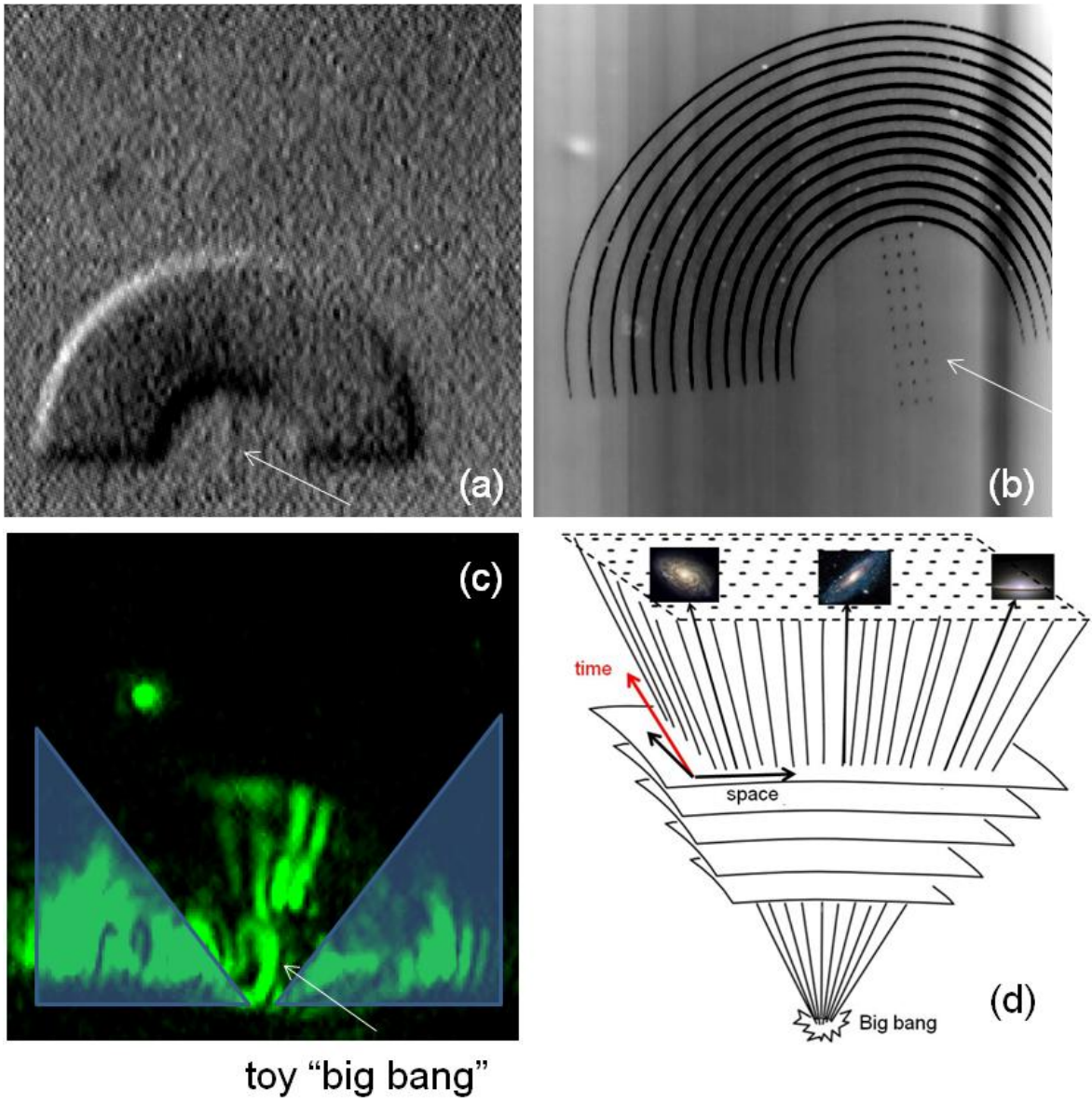

toy "big bang"

Fig.3

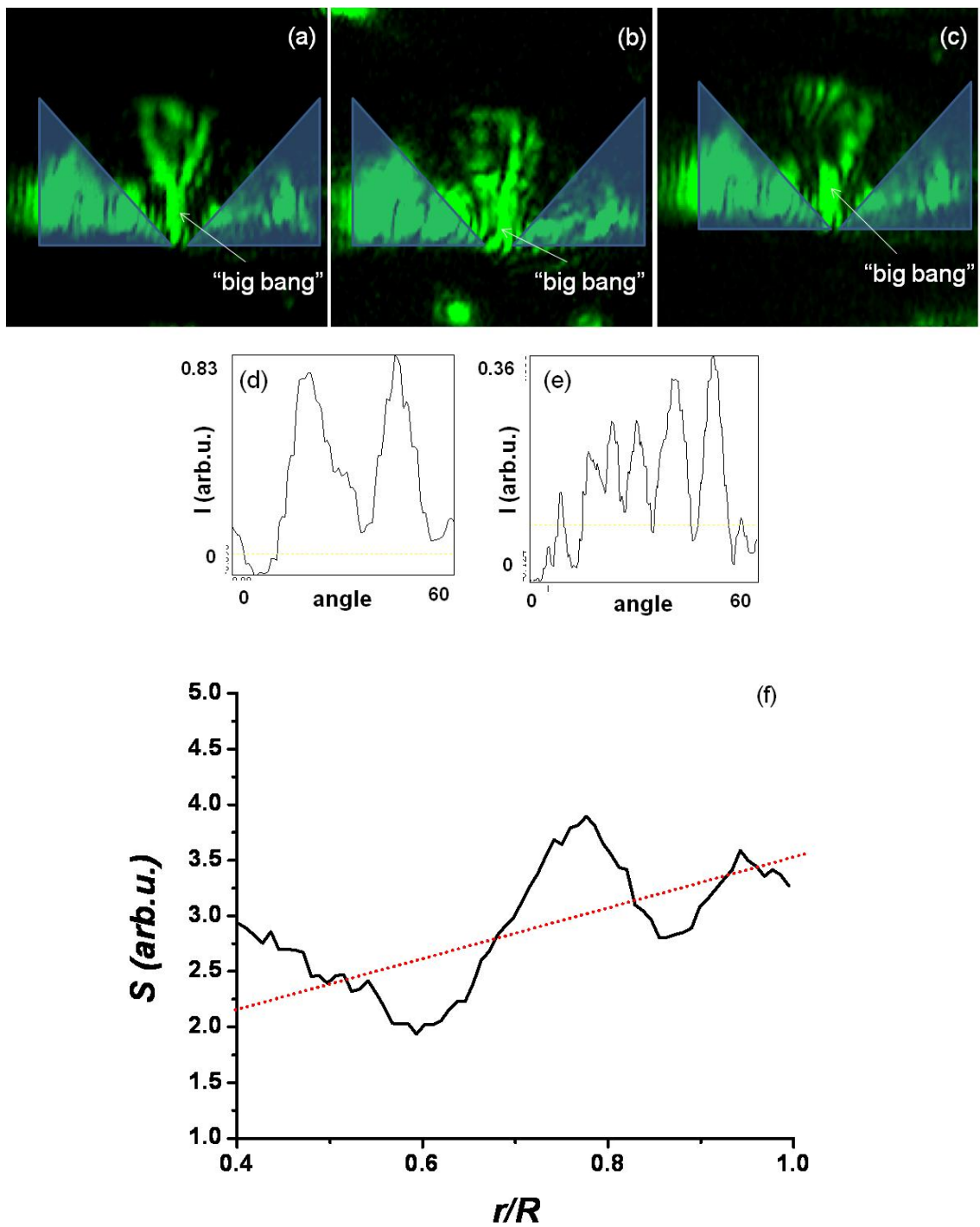

Fig.4